\documentclass[12pt]{iopart}

\usepackage{graphicx}
\usepackage{epstopdf}
  
\begin{document}
\title[Optical fiber tsunamis]{Optical tsunamis: shoaling of shallow water rogue waves in nonlinear fibers with normal dispersion}

\author{Stefan Wabnitz$^1$}

\address{$^1$ Dipartimento di Ingegneria dell'Informazione,  Universit\`a degli Studi di Brescia, via Branze 38, 25123, Brescia, Italy}
\ead{stefano.wabnitz@ing.unibs.it}

\begin{abstract}
In analogy with ocean waves running up towards the beach, shoaling of prechirped optical pulses may occur in the normal group-velocity dispersion regime of optical fibers. We present exact Riemann wave solutions of the optical shallow water equations and show that they agree remarkably well with the numerical solutions of the nonlinear Schr\"odinger equation, at least up to the point where a vertical pulse front develops. We also reveal that extreme wave events or optical tsunamis may be generated in dispersion tapered fibers in the presence of higher-order dispersion.
\end{abstract}
\pacs{42.65.Tg, 42.65.Sf, 42.81.Dp, 42.65.Re, 47.35.Fg}
\vspace{2pc}
\noindent{\it Keywords}: Nonlinear optics, Fluid dynamics, Optical fibers, Optical rogue waves. 
\maketitle

\section{Introduction}
\label{sec:intro}
Extreme or rogue waves have received a great deal of attention recently for their emergence in a variety of applications, ranging from fluid dynamics and oceanography to plasma physics, Bose-Einstein condensation (BEC) and nonlinear optics \cite{dhyste2008}-\cite{akhmediev2010sp}. The most popular manifestation of a rogue wave has so far been described as the sudden build-up and subsequent rapid disapperance in the open sea of an isolated giant wave, whose height and steepness are much larger than the corresponding average values of other waves in the ocean. A different manifestation of rogue waves with the potential for large scale damages also occurs as a result of wave propagation in shallow waters, a framework which describes the run-up of tsunamis towards the coast \cite{didenkulova}. Moreover, it is also known that the crossing of waters propagating in different directions or with opposite velocities may lead to the formation of high-elevation and steep humps of water or sneaker waves, which may cause severe disruptions along the coastine and river flooding \cite{soomere}. The same phenomena also occur in the run-down of avalanches falling from a mountain channel or a glacier \cite{zahibo}.

In deep waters, the dynamics of rogue wave formation may be described by the well-known one-dimensional
nonlinear Schr\"odinger equation (NLSE). As such, the generation of rogue waves has been closely associated so far with the presence of continuous wave (CW) breaking or modulation instability (MI) \cite{benjamin}, which occurs in the so-called self-focusing or anomalous group-velocity dispersion (GVD) regime. Thanks to the integrability of the NLSE, the nonlinear development of the MI originates families of exact solutions such as the Akhmediev breathers \cite{akhmediev86}. Although generally periodic both in the evolution variable (e.g., distance) and the transverse dimension (e.g., time), such nonlinear wave families may also include strictly spatio-temporally localized waves: consider for example the Peregrine soliton \cite{peregrine83}.

Given the widespread applicability of the NLSE to fields of physics other than oceanography, it has been possible to predict and experimentally observe rogue wave phenomena in different contexts such as nonlinear optical fibers. For example, the temporal statistics of optical supercontinuum generation has revealed the emergence of extreme solitary wave emissions \cite{solli2007}. In addition, optical fibers have provided the test-bed for the the first clear experimental observation of Peregrine solitons by inducing the MI of a pump laser by means of an additional seed laser\cite{kibler2010}. 

On the other hand, the appearance of rogue waves in shallow waters is a field which is still the emerging stage \cite{didenkulova}. As such, it does not appear to have received much attention in other physical contexts besides hydrodynamics. A recent study has however pointed out that extreme waves may be also be generated in optical fibers in the normal GVD regime of pulse propagation, where MI is absent \cite{wab2013}. In this famework the model linking hydrodynamics with nonlinear optics is provided by the semiclassical approximation to the NLSE \cite{yuji95}, which is known as the nonlinear shallow water equation (NSWE) \cite{whitham} or the Saint-Venant equation \cite{stvenant}. Although in the normal GVD regime a CW is modulationally stable, extreme waves may still be generated in the presence of a suitable temporal pre-chirp or phase modulation \cite{wab2013},\cite{yuji99}-\cite{biondini06}, which is analogous to a nonuniform velocity distribution of the propagating water waves, eventually leading to tsunamis. As well known, the conditions of shallow water (as opposed to deep water) propagation applies in oceanography whenever the wavelength of the wave is much longer than the depth of the water, such as it happens for tsunamis even at large distances from the coast, given their typical wavelengths of the order of thousands of km. On the other hand in nonlinear optics, the applicability of the NSWE requires the characteristic dispersion distance to be much longer than the nonlinear distance (small dispersion limit)\cite{yuji95}.

In this work we show that, in analogy with the commonly experienced case of ocean waves as they run-up to the beach, the shoaling of properly pre-chirped optical pulses may also occur in the normal dispersion regime of optical fibers. Therefore we shall consider the propagation of special, input temporally pre-chirped optical pulses with different power profiles. These pulses represent nonlinear invariant solutions of the NSWE (Riemann waves). For such type of chirped pulses, we obtain exact solutions of the optical NSWE, and demonstrate their good agreement with numerical solutions of the NLSE, at least up to the point where a vertical pulse edge or front develops in the power profile. In addition, we also present simulations which reveal that third-order dispersion (TOD) leads to the occurrence of extreme waves or optical tsunamis whenever a dispersion tapered fiber is used, in analogy with the dramatic run-up and wave height amplification of a tsunami as the coast is approached and the water depth is progressively reduced.

In section \ref{sec:eqs} we present the NSWE and its formulation in Riemann invariant form, which leads to the definition of the nonlinear Riemann waves. Exact solutions for these temporally chirped optical Riemann pulses with different input power temporal profiles, such as the parabolic, Gaussian or hyperbolic secant profile, are presented in section \ref{sec:nrw}, and critically compared with the numerical solutions of the NLSE. The analysis of section \ref{sec:nrw} reveals the emergence of wave-breaking free optical wave shoaling, that is the formation of a vertical edge for the propagating pulse, without any appearance of high frequency temporal oscillations or shock phenomena. Finally, as discussed in section \ref{sec:rw}, we show that whenever the balance of GVD and TOD dynamically evolves along the fiber, such as it occurs in dispersion varying or tapered fibers, significant temporal compression and peak power amplification may result in the form of a high-intensity spatio-temporally localized flash of light propagating entirely in the normal GVD regime. 

\section{Basic equations}
\label{sec:eqs}

As it is well known, the propagation of short light pulses in optical fibers may be described in terms of the NLSE

\begin{equation}
	\nonumber i\frac{\partial Q}{\partial z}-\frac{\beta_2}{2}\frac{\partial^2 Q}{\partial t^2}-i\frac{\beta_3}{6}\frac{\partial^3 Q}{\partial t^3}+\gamma|Q|^2Q=\alpha_f t Q .\label{nls1}
\end{equation}

\noindent Here $z$ and $t$ denote the distance and retarded time (in the frame travelling at the carrier frequency group-velocity) coordinates; $\beta_2$, $\beta_3$ and $\gamma$ represent GVD, TOD and the nonlinear coefficient, respectively; $Q$ is the field envelope. Moreover, $\alpha_f$ denotes the rate of the carrier frequency shift with distance. A continuous frequency shift may be introduced in a fiber ring laser by means of an acousto-optic filter \cite{yuji94}. Otherwise, the frequency shifting term may be introduced to represent, in combination with the TOD term, propagation in a dispersion tapered optical fiber, where the local dispersion varies along the propagation distance \cite{latkin07}-\cite{wab08}. In dimensionless units, and in the normal GVD regime (i.e., $\beta_2>0$), Eq.(\ref{nls1}) reads as

\begin{equation}
	\nonumber i\frac{\partial q}{\partial Z}-\frac{\beta^2}{2}\frac{\partial^2 q}{\partial T^2}-i\frac{\widehat{\beta}}{6}\frac{\partial^3 q}{\partial T^3}+|q|^2q=\hat{\alpha} T q.\label{nls2}
\end{equation}

\noindent Here $T=t/t_0$, $Z=z\gamma P_0=z/L_{NL}$, $\beta^2=\beta_2/(t_0^2\gamma P_0)\equiv L_{NL}/L_D$, where $L_{NL}$ and $L_D$ are the nonlinear and dispersion lengths, respectively, $\widehat{\beta}=\beta_3/(t_0^3\gamma P_0)$, $q=Q/\sqrt{P_0}$, and $\hat{\alpha}=\alpha_f t_0/(\gamma P_0)$; $t_0$ and $P_0$ are arbitrary time and power units. Eq.(\ref{nls2}) can be expressed in terms of the real variables $\rho$ and $u$ which denote the field dimensionless power and instantaneous frequency (or chirp)

\begin{equation}
	\nonumber q(T,Z)=\sqrt{\rho(T,Z)}\exp\left[-\frac{i}{\beta}\int_{-\infty}^{T}u(T',Z)dT'\right].\label{ansatz}
\end{equation}
 
By ignoring higher order time derivatives in the resulting equations (which is justified for relatively small values of $\beta$) as well as TOD, one obtains from the NLSE the hydrodynamic NSWEs \cite{yuji95}-\cite{whitham}

\begin{equation}\label{sweq}
\left \{ \begin{array} {l}
\displaystyle\frac{\partial \rho}{\partial Z'} + \frac{\partial \left(\rho u\right)}{\partial T} = 0 \\
\\
\displaystyle\frac{\partial u}{\partial Z'} + u \frac{\partial u }{\partial T} + \frac{\partial \rho}{\partial T} = \alpha,
\end{array} \right .
\end{equation}

\noindent where $Z'=\beta Z$ and $\alpha=\hat{\alpha}/\beta$. In hydrodynamics, Eqs.(\ref{sweq}) describe the motion of a surface wave in shallow water, i.e., a wave whose wavelength is much larger than the water depth. In this context, 
$\rho$ and $u$ represent the water depth and its velocity, respectively. Moreover $\alpha=dh/dT$, where $h(T)$ is the unperturbed water depth along the channel or beach axis \cite{didenkulova}. Therefore a constant $\alpha$ represents tsunami run-up towards a beach with uniform slope. 

Solutions of Eqs.(\ref{sweq}) may be found by making use of the Riemann invariants 

\begin{equation}\label{riemann}
I_{\pm}=u \pm 2\sqrt{\rho}-\alpha T
\end{equation}

\noindent so that Eqs.(\ref{sweq}) read as

\begin{equation}\label{invariant}
\frac{\partial  I_{\pm}}{\partial Z'} + c_{\pm} \frac{\partial  I_{\pm}}{\partial T} =0
\end{equation}

\noindent where

\begin{equation}\label{speed}
c_{\pm} =\frac{3}{4} I_{\pm}+\frac{1}{4} I_{\mp}+\alpha T
\end{equation}

If we set $u=\alpha T + \nu$, so that

\begin{equation}\label{change}
\left \{ \begin{array} {l}
\displaystyle\nu=\frac{I_+ + I_-}{2} \\
\\
\displaystyle\rho=\frac{\left(I_+ - I_-\right)^2}{16},
\end{array} \right .
\end{equation}

\noindent use the accelerated reference frame $T'=T-\alpha Z'^2/2$, $Z'=Z'$, and drop primes for simplicity,
 Eqs.(\ref{invariant}) read as

\begin{equation}\label{invariant1}
\frac{\partial  J_{\pm}}{\partial Z} + C_{\pm} \frac{\partial  J_{\pm}}{\partial T} =0
\end{equation}

\noindent where

\begin{equation}\label{change1}
\left \{ \begin{array} {l}
\displaystyle J_{\pm}=\nu \pm 2 \sqrt{\rho}\\
\\
\displaystyle C_{\pm}=\frac{3}{4} J_{\pm}+\frac{1}{4} J_{\mp}
\end{array} \right .
\end{equation}

\noindent For a temporally localized input optical waveform such as a chirp-free square pulse, i.e., with $\rho(T,Z=0)=\rho_0$ for $\left|T\right|\leq T_0$ and $\rho(T,Z=0)=0$ otherwise, Eq.(\ref{sweq}) may be analytically solved up to the point $Z=T_0/\sqrt{\rho_0}$ in terms of the well-known Ritter dam-break solution \cite{yuji95},\cite{ritter}. On the other hand, henceforth we shall restrict our attention to the propagation of special pre-chirped optical pulses, as discussed in the next section.  

\section{Nonlinear Riemann waves}
\label{sec:nrw}

A particular solution of Eqs.(\ref{invariant1}), known as the Riemann wave, can be obtained if we set $J_{-}=0$, so that $\nu=2\sqrt{\rho}$. Therefore Eqs.(\ref{invariant1}) reduce to

\begin{equation}\label{rw}
\frac{\partial  \Gamma}{\partial Z} + \Gamma \frac{\partial  \Gamma}{\partial T} =0
\end{equation}

\noindent where $\Gamma\equiv 3J_{+}/4=3\nu/2=3\sqrt{\rho}$. The solution of Eq.(\ref{rw}) reads as $\Gamma(T,Z)=\Gamma_0(T-\Gamma Z)$, where $\Gamma_0(T)=\Gamma(T,Z=0)$. In practice it is easier to consider a given intial power profile for the optical pulse. From the first of Eqs.(\ref{sweq}), one also obtains for the pulse power the following equation

\begin{equation}\label{eqrho}
\frac{\partial  \rho}{\partial Z} + 3\sqrt{\rho} \frac{\partial  \rho}{\partial T} =0
\end{equation}

\noindent The solution of Eq.(\ref{eqrho}) corresponding to the initial power profile $\rho(T,Z=0)=\rho_0 P(T/T_0)$ can be written in implicit form as \cite{rudenko}

\begin{equation}\label{powero}
\rho(T,Z)=\rho_0 P\left[\left(T-3Z\sqrt{\rho(T,Z)}\right)/T_0\right]
\end{equation}

\noindent or, in more compact notation, as

\begin{equation}\label{powersh}
P^{-1}(p)=\tau -a \zeta \sqrt{p}
\end{equation}

\noindent where we defined $p=\rho/\rho_0$, $\tau=T/T_0$, $a=3\sqrt{2\rho_0/T_0}$, and $\zeta=Z/\sqrt{2T_0}$. In the following subsection, we will apply Eq.(\ref{powersh}) to exactly solve some specific cases of interest in nonlinear optics, namely a pulse with an initial power profile of either parabolic, Gaussian or hyperbolic secant shape. 

\subsection{Parabolic pulse}

\begin{figure}[ht]
\centering
\includegraphics[width=10cm]{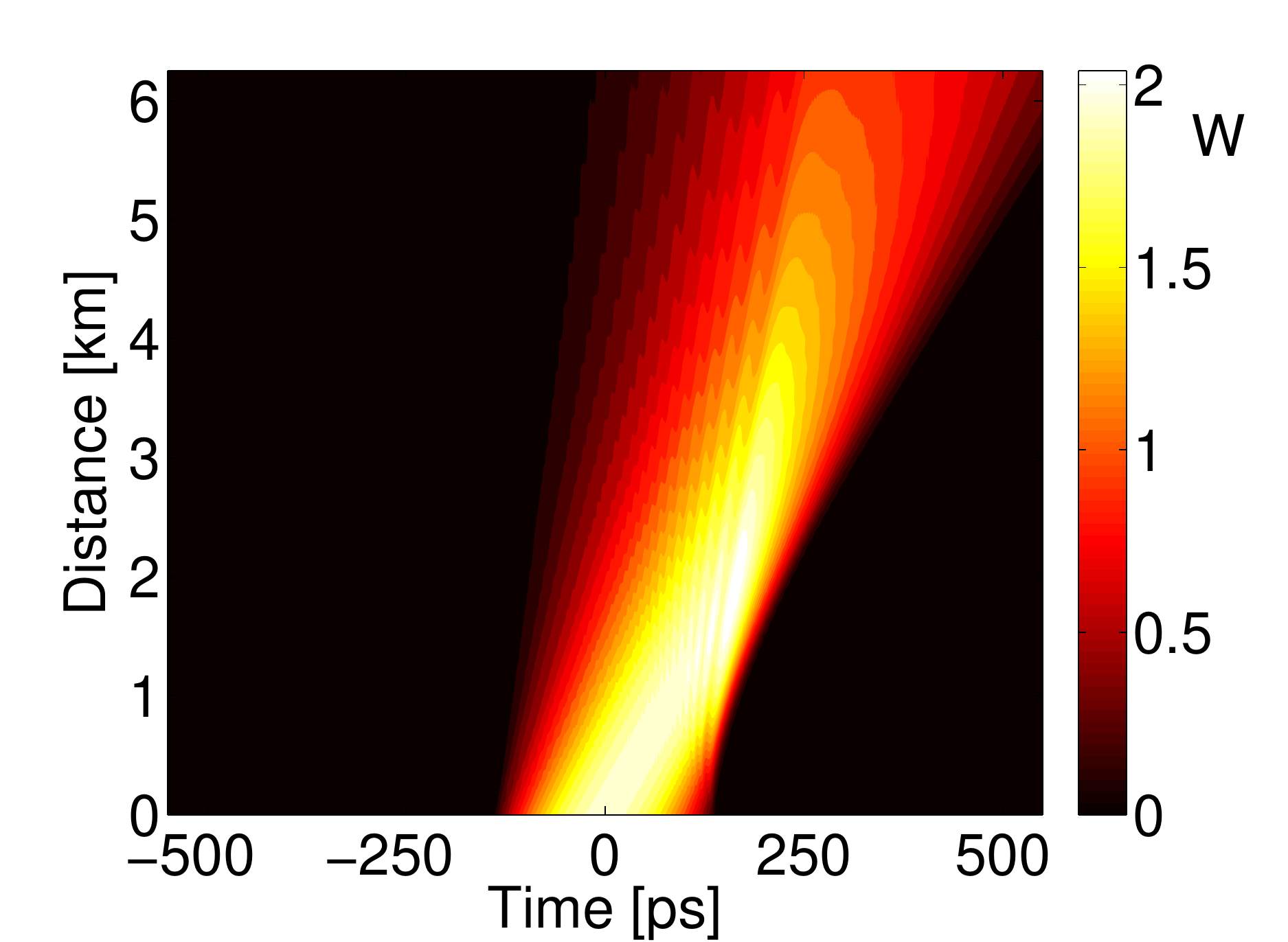}
\caption{Contour plot of pulse power vs. fiber length by solving Eq.(\ref{nls1}) with an initial parabolic power profile.}
\label{fig1}
\end{figure}

Consider first the interesting case of a pulse with initial (i.e., at $\zeta=0$) parabolic power profile of finite duration $2T_0$ (or parabolic cap), that is

\begin{equation}\label{parabola}
	p(\tau,\zeta=0)=\left\{\begin{array} {l}
\displaystyle 1-\tau^2 \: for \:\left|\tau \right|<1,\\ 
\\
\displaystyle 0 \: for \:\left|\tau \right|\geq 1\end{array}\right.
\end{equation}

\noindent One easily obtains from Eq.(\ref{powersh}) the solution for the power profile at any $\zeta$ as 

\begin{equation}\label{parabolaev}
	\sqrt{p(\tau,\zeta)}=\left\{\begin{array} {l}
\displaystyle \frac{-B + \sqrt{\Delta}}{A} \: for \:\left|\tau -a\zeta \sqrt{p} \right|<1,\\ 
\\
\displaystyle 0 \:for \:\left|\tau -a\zeta \sqrt{p}\right|\geq 1\end{array}\right. 
\end{equation}

\noindent where 

\begin{equation}\label{defs}
A=1+a^2\zeta^2,\: B=-a\zeta\tau, \: C=-1+\tau^2, \: \Delta=B^2-A C.
\end{equation}

\noindent and the inequalities in Eq.(\ref{parabolaev}) immediately result by imposing that the pulse power $p$ remains equal or larger than zero in Eq.(\ref{powersh}). The corresponding temporal chirp profile is always provided by the relationship $\nu=2\sqrt{\rho}$.

\begin{figure}[ht]
\centering
\includegraphics[width=10cm]{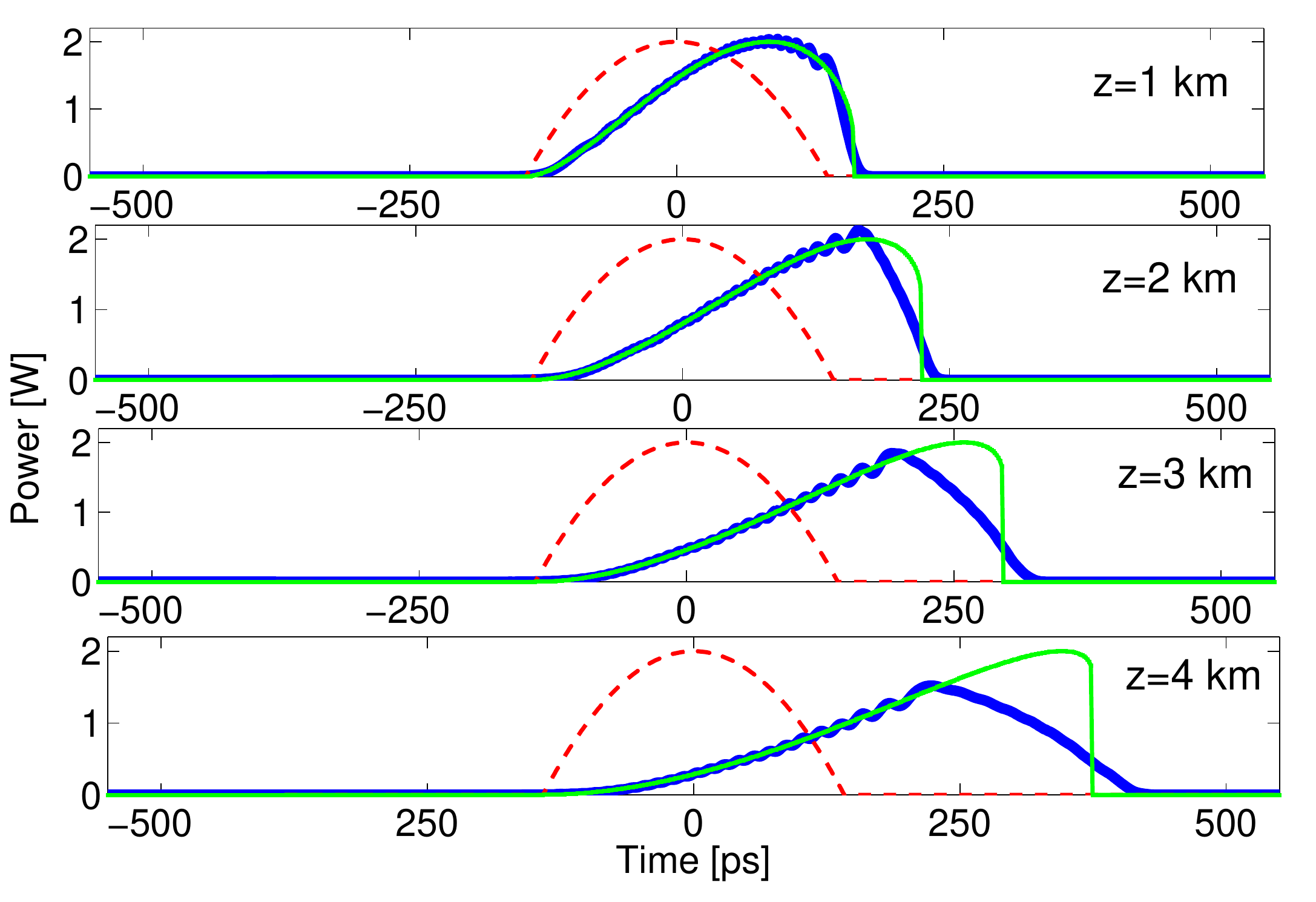}
\caption{Pulse power temporal profiles at the output of different DCF lengths: blue solid thick curves indicate numerical solution of the NLSE; green solid thin curves show the exact solution of the NSWEs; the red dashed curve shows the input parabolic power profile.}
\label{fig2}
\end{figure}

It is interesting to compare the analytical solution (\ref{parabolaev}) of the NSWEs (\ref{sweq}) with the numerical solution of the original NLSE (\ref{nls1}). Let us consider the specific case of a dispersion-compensating fiber (DCF) with normal GVD coefficient $D=-100 ps/(nm\cdot km)$ (or $\beta_2=127\: ps^2/km$ at $1550\: nm$), and the nonlinear coefficient $\gamma =3.2 W^{-1}km^{-1}$ in Eq.(\ref{nls1}). 

In Fig.\ref{fig1} we illustrate the propagation of a nonlinear Riemann wave with initial parabolic power profile by showing the contour plot of the pulse power as a function of the distance along the DCF, as it is computed by solving the original NLSE (\ref{nls1}). Here the input peak pulse power is equal to $P=2\: W$, and the parabolic pulse full-width at half maximum is $T_{fwhm} =200\: ps$. Fig.\ref{fig1} shows that the pre-chirp distribution leads to a slowing down of the pulse. Moreover, the pulse power profile develops a strongly asymmetric shape: its trailing edge acquires a vertical slope at about $z=1\: km$. For longer distances both trailing and leading edges of the pulse exhibit a nearly linear decrease with time, albeit with different slopes. The details of the evolution of the pulse power profile with fiber length is clearly illustrated in Fig.\ref{fig2}: here we compare the numerical solution of the NLSE (\ref{nls1}) (blue solid thick curves) with the exact solution of the NSWEs (green solid thin curves). In Fig.\ref{fig2} we also show the input parabolic power profile (red dashed curves). 

As it can be seen in Fig.\ref{fig2}, there is an excellent agreement between the numerical solution of the NLSE and the exact solution (\ref{parabolaev}) up to the point (i.e., $z=1\: km$) where a wavefront with vertical slope develops at the trailing edge of the pulse. For longer distances, the agreement between the analytical NSWE solution and the numerical NLSE solution remains for the entire leading edge of the pulse. On the other hand the vertical trailing edge is preserved in the analytical solution of the NSWEs, whereas the trailing edge shows a smoothened decay in the solution of the NLSE. The optical pulse deformation which is seen in Figs.\ref{fig1}-\ref{fig2} is analogous to the shoaling of a shallow water wave when it approaches to the beach. Therefore we may name the point where the vertical pulse edge develops as the shoaling point. It is remarkable that no wave breaking occurs in spite of the occurrence of a vertical slope in the pulse trailing edge: the NLSE solutions exhibit a self-regularisation behavior.  

\begin{figure}[ht]
\centering
\includegraphics[width=10cm]{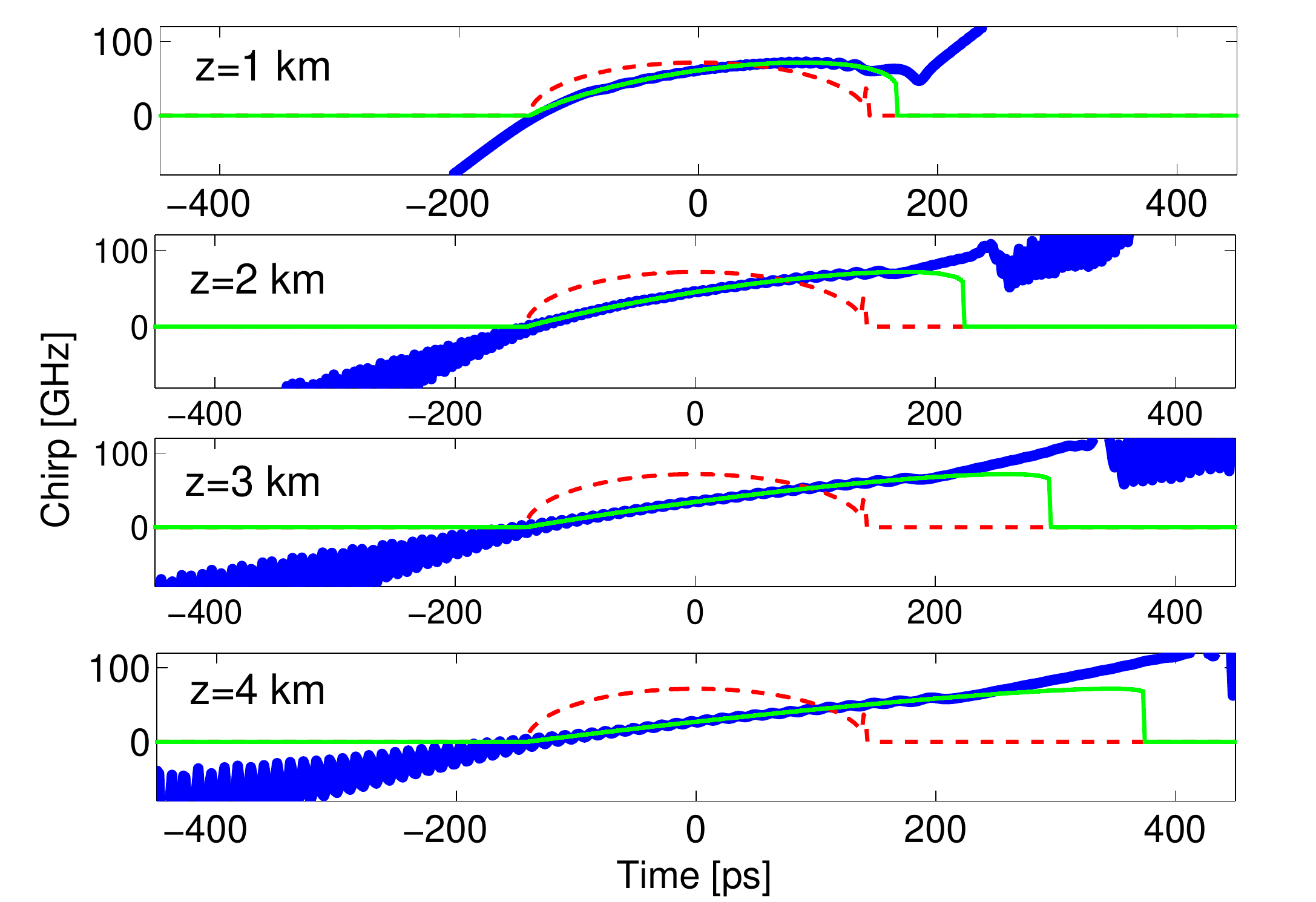}
\caption{Pulse chirp temporal profiles at the output of different DCF lengths: blue solid thick curves indicate numerical solution of the NLSE; green solid thin curves show the exact solution of the NSWEs; red dashed curves show the input parabolic power profile.}
\label{fig3}
\end{figure}

In Fig.\ref{fig3} we display the evolution of the chirp profiles with distance, corresponding to the power profiles of Fig.\ref{fig2}. Here the red dashed curves indicate the initial chirp profile of the parabolic Riemann pulse. As can be seen,  the analytical solutions for the chirp, as they are obtained either from the NSWE or the NLSE, are in excellent agreement with each other up the shoaling point. For longer distances, the agreement only remains for the (relatively longer) leading edge of the pulse.  

\subsection{Gaussian pulse}

\begin{figure}[ht]
\centering
\includegraphics[width=10cm]{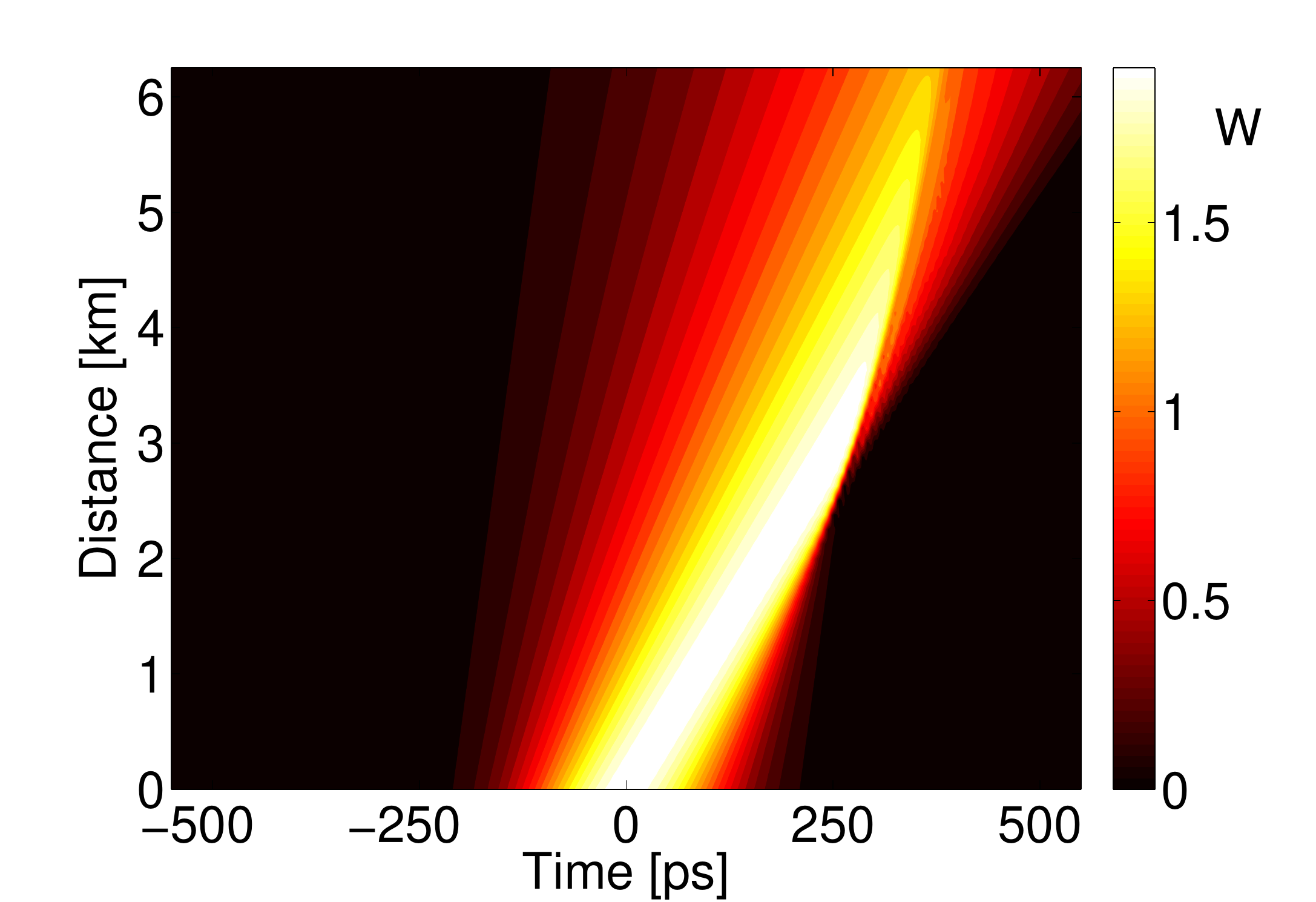}
\caption{Contour plot of pulse power vs. fiber length by solving Eq.(\ref{nls1}) with initial Gaussian power profile.}
\label{fig4}
\end{figure}

In the case of a nonlinear Riemann wave with an input Gaussian power profile, we may set $p(\tau,\zeta=0)=\exp{\left(-2\tau^2\right)}$, so that from Eq.(\ref{powersh}) we obtain the implicit equation for the pulse power

\begin{equation}\label{gauss}
p(\tau,\zeta)=\exp{\left[-2\left(\tau^2-2a\zeta \tau \sqrt{p}+a^2\zeta^2 p\right)\right]}
\end{equation}

\noindent which can be easily solved by an iterative procedure. In Fig.\ref{fig4} we display the corresponding contour plot of the pulse power vs. distance along the DCF, as it is obtained by solving the NLSE (\ref{nls1}). The input Gaussian peak pulse power is always equal to $P=2\: W$, and the pulse $T_{fwhm} =200\: ps$. Fig.\ref{fig4} shows that with a Gaussian pulse the shoaling point occurs at about $z=2.5\: km$. 

\begin{figure}[ht]
\centering
\includegraphics[width=10cm]{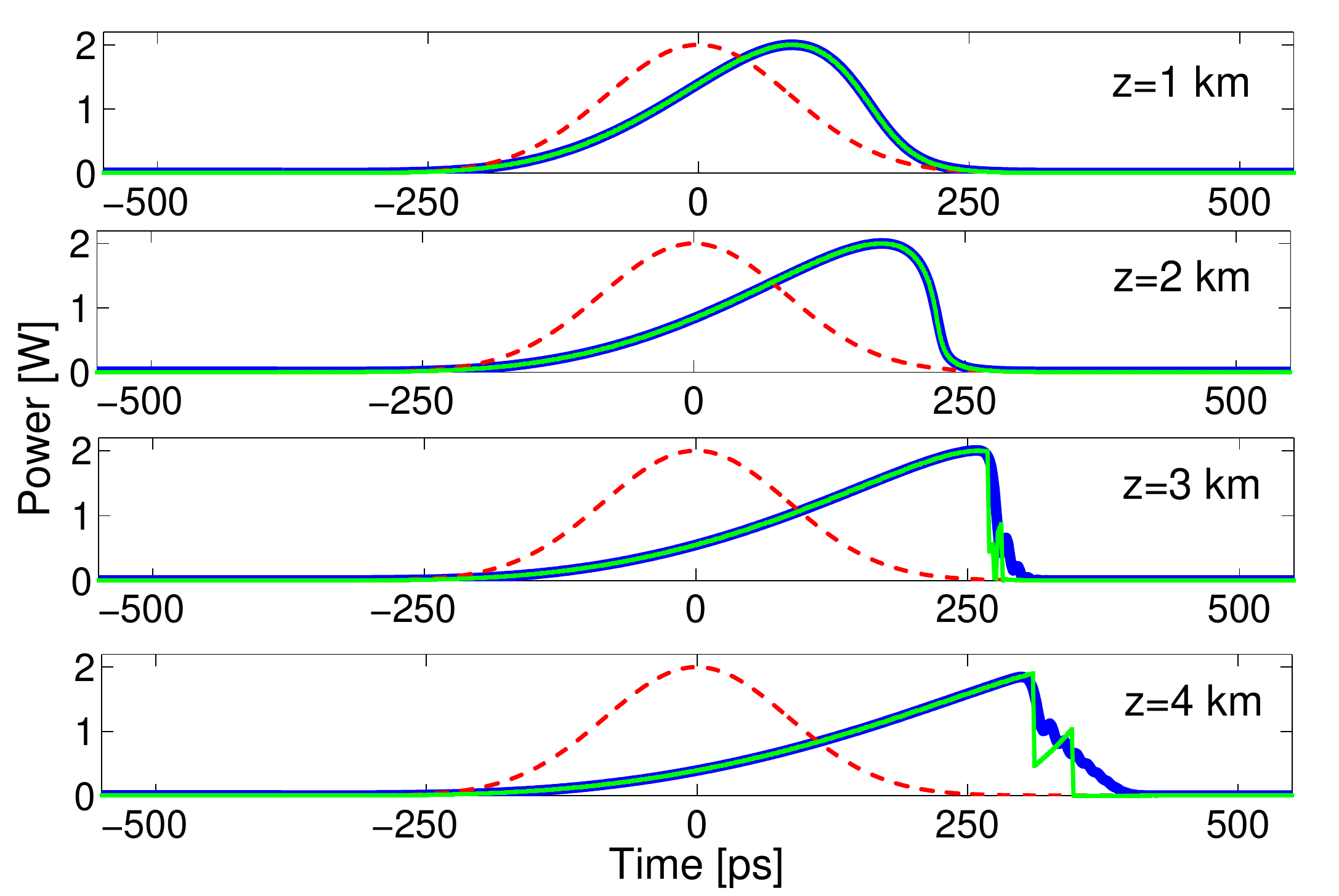}
\caption{Same as in Fig.\ref{fig2}, with initial Gaussian pulse power profile.}
\label{fig5}
\end{figure}

The comparison among the exact NSWE and numerical NLSE solutions for the pulse power profiles shown in Fig.\ref{fig5} reveals that their agreement is very good up to the shoaling point, that is $z\cong 3\: km$. For longer distances, the NLSE solution develops a tail in the trailing edge which is not captured by the solution (\ref{gauss}) of the NSWE (\ref{sweq}). Moreover, the fragmented green curve in the bottom panel of Fig.\ref{fig5} (i.e., for $z= 4\: km$) shows that the numerical solution of the transcendental equation (\ref{gauss}) is no longer accurate for points in time beyond the vertical trailing edge.

\begin{figure}[ht]
\centering
\includegraphics[width=10cm]{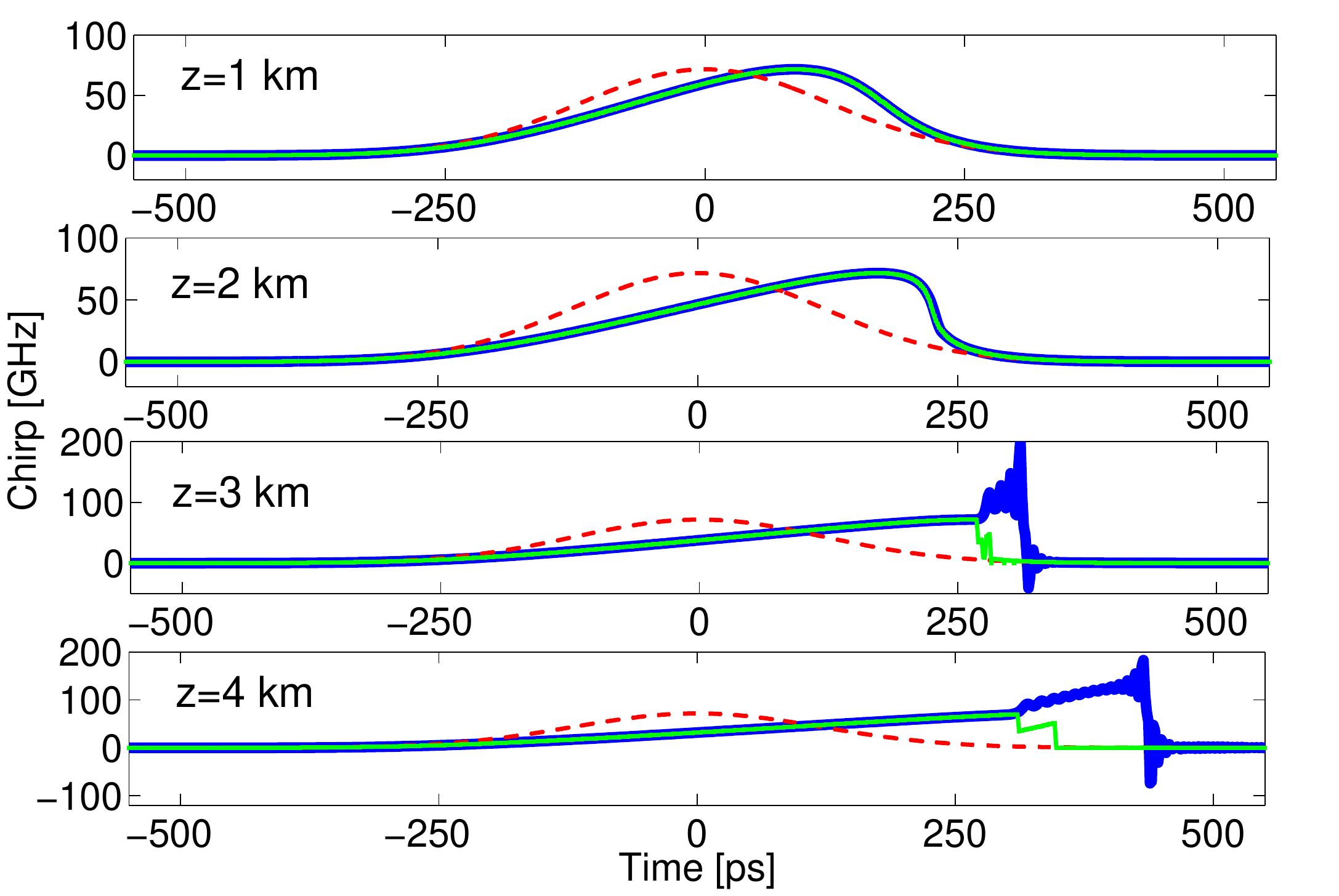}
\caption{Same as in Fig.\ref{fig3}, with initial Gaussian pulse power profile.}
\label{fig6}
\end{figure}

Moreover the comparison of the chirp profiles in Fig.\ref{fig6} shows that past the shoaling point the NLSE pulse acquires a temporally oscillating negative chirp in its leading edge, and a uniform positive chirp in its trailing edge. Such highly chirped regions in the pulse tails are not fully captured by the NSWE solution (\ref{gauss}).  

\subsection{Hyperbolic secant pulse}

\begin{figure}[ht]
\centering
\includegraphics[width=10cm]{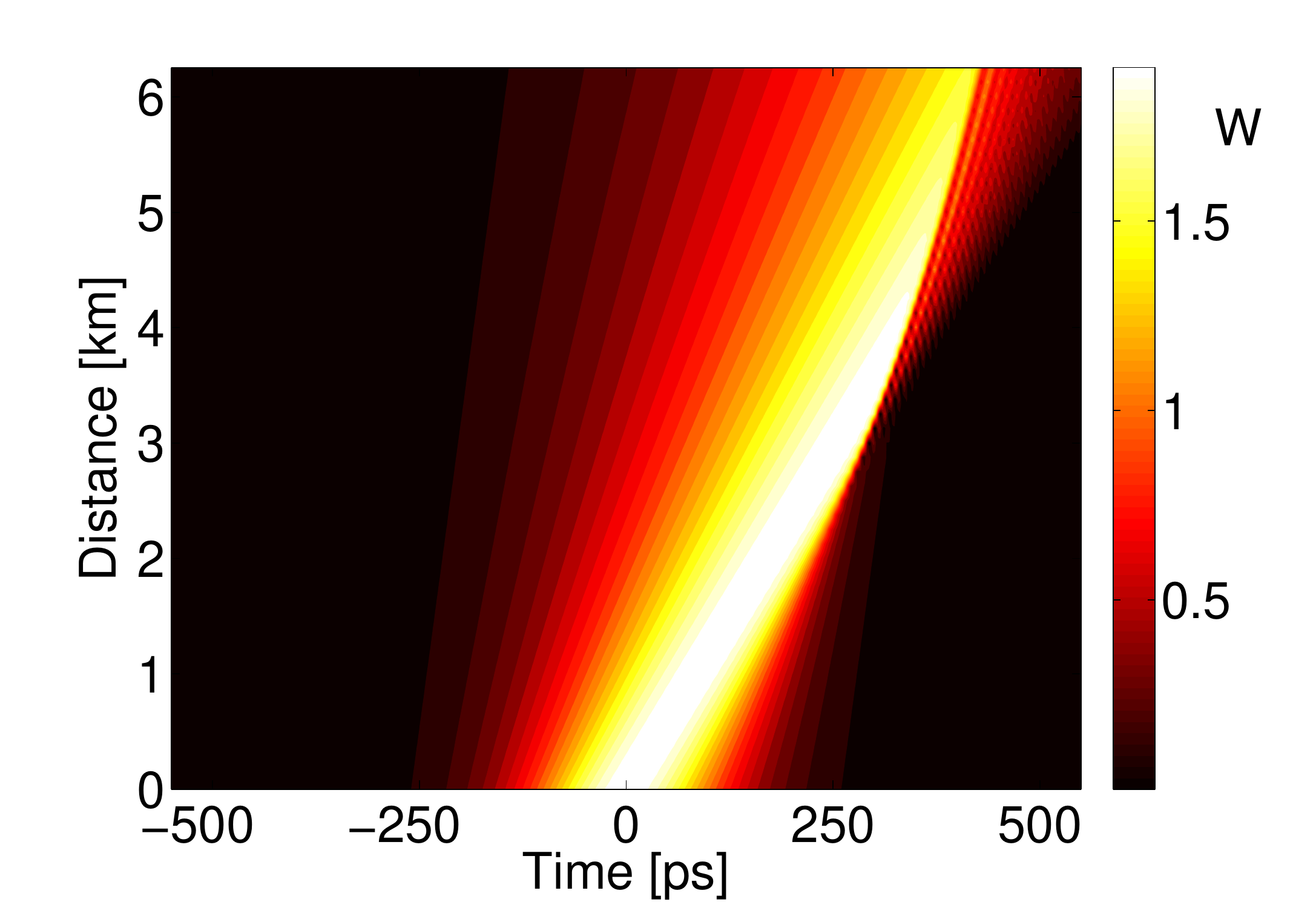}
\caption{Contour plot of pulse power vs. fiber length by solving Eq.(\ref{nls1}) with initial hyperbolic secant power profile.}
\label{fig7}
\end{figure}

In the case of an input hyperbolic secant pulse, we may set $p(\tau,\zeta=0)=sech^2(\tau)$ so that, from Eq.(\ref{powersh}) we obtain

\begin{equation}\label{sech}
p(\tau,\zeta)=sech^2\left(\tau-a\zeta\sqrt{p}\right)
\end{equation}

In Fig.\ref{fig7} we illustrate the contour plot of the pulse power vs. distance along the DCF from the solution of the NLSE (\ref{nls1}) for an input Riemann wave with initial hyperbolic secant power profile. The input peak pulse power is kept equal to $P=2\: W$, and $T_{fwhm} =200\: ps$. Fig.\ref{fig7} shows that with a hyperbolic secant pulse the shoaling point is shifted further up to about $z=3.5\: km$: otherwise the power evolution remains qualitatively very similar to the case of a Gaussian pulse which is shown in Fig.\ref{fig4}.  

\begin{figure}[ht]
\centering
\includegraphics[width=10cm]{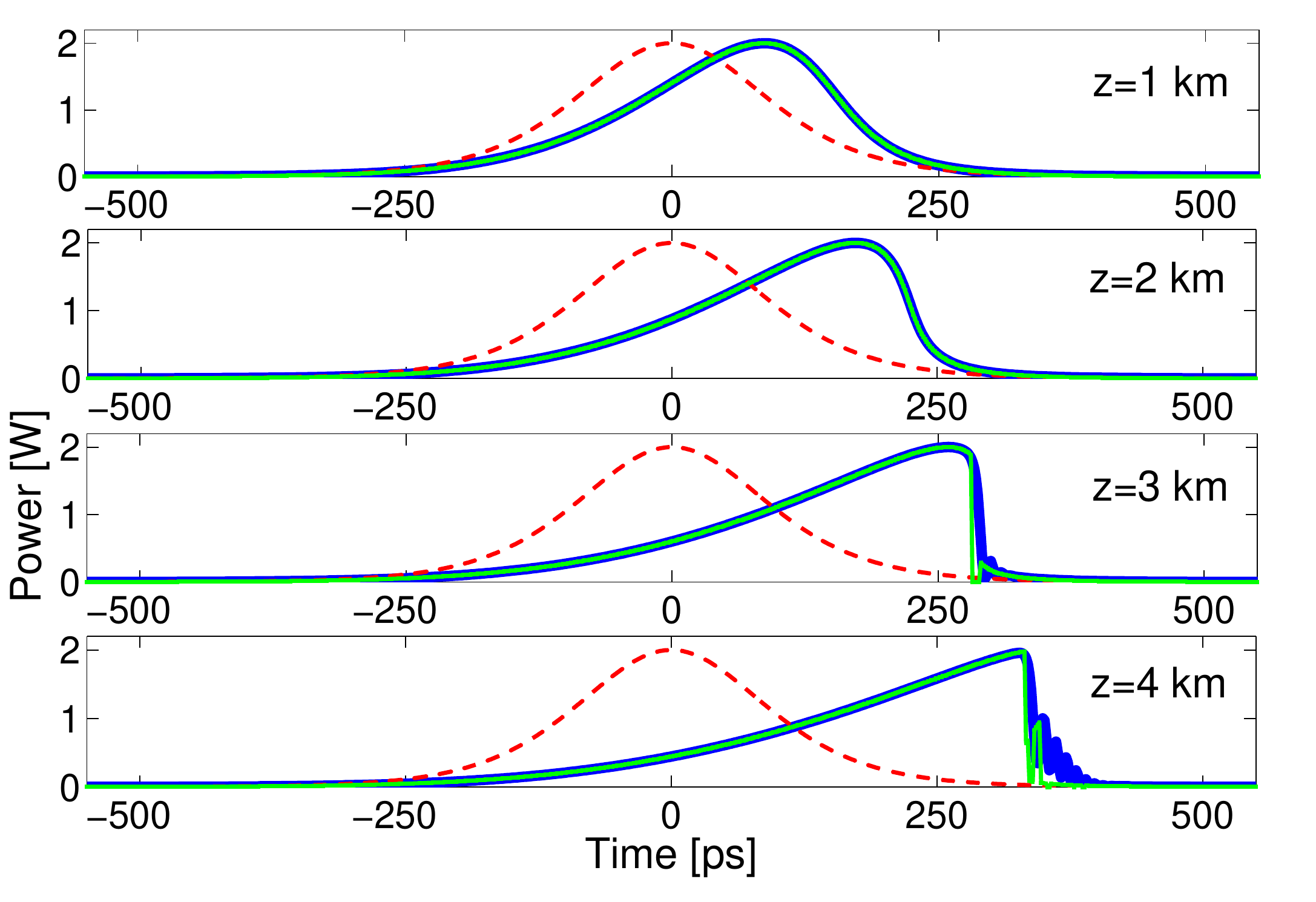}
\caption{Same as in Fig.\ref{fig2}, with initial hyperbolic secant pulse power profile.}
\label{fig8}
\end{figure}

Fig.\ref{fig8} illustrates the details of the pulse power with an initial hyperbolic secant profile for the exact NSWE and numerical NLSE solutions. Again, there is an excellent agreement up to the shoaling point, or $z\cong 4\: km$. 

\begin{figure}[ht]
\centering
\includegraphics[width=10cm]{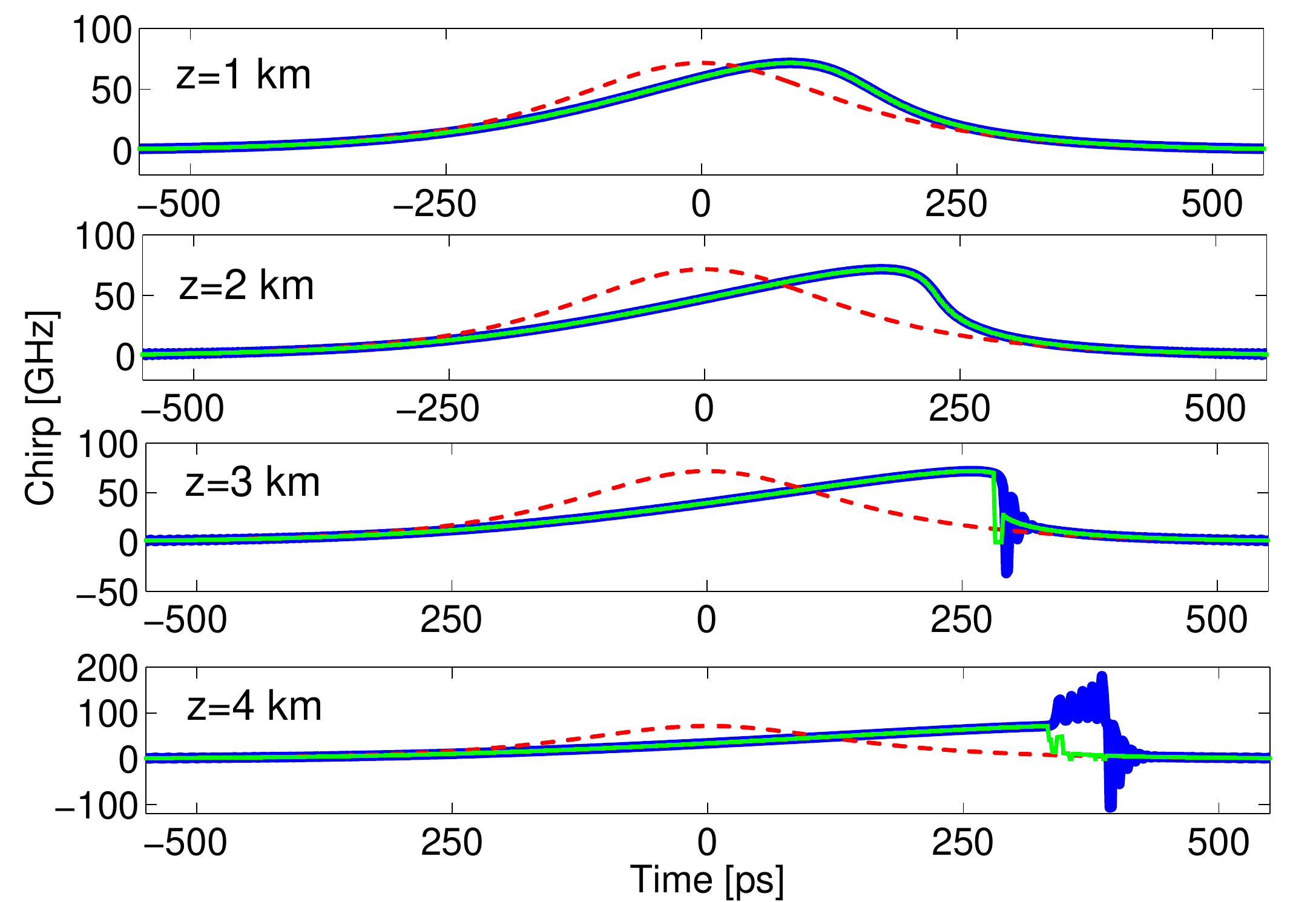}
\caption{Same as in Fig.\ref{fig3}, with initial hyperbolic secant pulse power profile.}
\label{fig9}
\end{figure}

On the other hand, the comparison of the chirp profiles which is illustrated in Fig.\ref{fig9} shows that past the shoaling point the NLSE pulse acquires a strong, temporally oscillating positive chirp in its trailing edge. Such highly chirped region corresponds to the oscillating tail in Fig.\ref{fig8}, which is not reproduced by the NSWE solution (\ref{sech}).

\section{Rogue waves}
\label{sec:rw}

In the previous section \ref{sec:nrw} we have shown that properly pre-chirped optical pulses with different power profiles experience the wave-breaking free formation of a vertical edge, which is analogous to the shoaling of shallow water waves. On the other hand, in spite of the strong temporal deformation, the value of the peak power remained nearly unchanged and close to the input value in the course of propagation. 
In this section we will demonstrate by numerical simulations that extreme wave formation, that is the emergence of temporally compressed, high intensity and transient pulses, is possible in optical fibers with the inclusion of frequency shifting and TOD. This is equivalent to the case of a dispersion varying optical fiber, where the local value of the dispersion changes owing to the variation of the fiber diameter \cite{latkin07}-\cite{wab08}.  Let us consider next the impact of fiber TOD on the propagation of nonlinear Riemann waves in optical fibers: the corresponding modified NSWE reads as \cite{yuji96}

\begin{equation}\label{sweqtod}
\left \{ \begin{array} {l}
\displaystyle\frac{\partial \rho}{\partial Z'} + \left(1+\bar{\beta}u\right)u\frac{\partial \rho}{\partial T} + \left(1+2\bar{\beta}u\right)\rho\frac{\partial u}{\partial T}= 0 \\
\\
\displaystyle\frac{\partial u}{\partial Z'} + \left(1+\bar{\beta}u\right)u \frac{\partial u }{\partial T} + \frac{\partial \rho}{\partial T} = \alpha,
\end{array} \right .
\end{equation}

\noindent where $\bar{\beta}=\widehat{\beta}/(2\beta^3)$. 

\begin{figure}[ht]
\centering
\includegraphics[width=10cm]{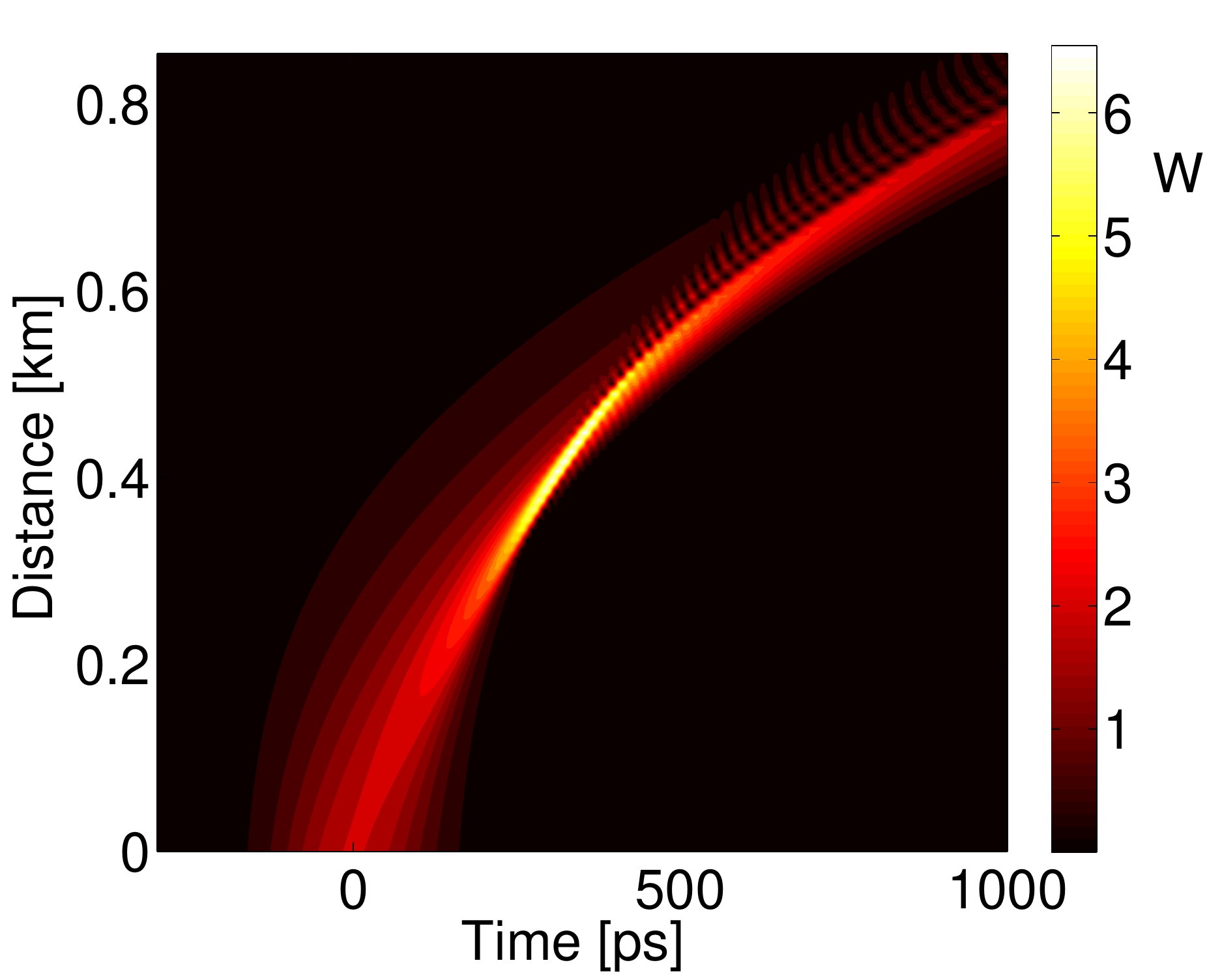}
\caption{Same as in Fig.\ref{fig6}, with negative third-order dispersion}
\label{fig10}
\end{figure}

In the following, we restrict our attention to investigating the possibility of the emergence of spatio-temporally localized flashes or rogue-wave type of solutions under the action of nonlinearity, dispersion, frequency-shifting and TOD by numerically solving the original NLSE (\ref{nls1}). For simplicity, we will only consider the case of an initial Riemann pulse with a Gaussian power profile, for the same power and temporal duration values as in the previous section. 

Fig.\ref{fig10} shows the contour plot of the pulse power evolution along the DCF whenever the frequency up-shift rate is equal $\alpha_f=256 \: MHz/m$ and the TOD equal to $\beta_3=-2 \: ps^3/m$. Note that the zero-dispersion point is up-shifted by $31.4 \: GHz$ with respect to the carrier frequency of the input pulse. The direction of frequency shifting is such that the fiber GVD is reduced by TOD as the pulse propagates in the fiber, so that the zero-dispersion point is crossed and eventually the pulse moves into the anomalous dispersion regime.

In Fig.\ref{fig10} and in the rest of this section, we will display the pulse evolution in the original temporal frame and not in the accelerated frame as it was done in the previous section. As we shall see, using the original reference frame permits a good temporal separation of the various pulse power profiles which are obtained at different distances.
As it can be seen by comparing Fig.\ref{fig10} with Fig.\ref{fig4}, the relatively large TOD value has the effect of anticipating the shoaling point (or vertical trailing edge) to $z=0.3\: km$  (from $z=2.5\: km$ in the absence of TOD). Moreover, a significant temporal compression and three-fold peak power amplification (from $2\: W$ to $6\: W$) occurs in correspondence of the shoaling point.   
   
\begin{figure}[ht]
\centering
\includegraphics[width=10cm]{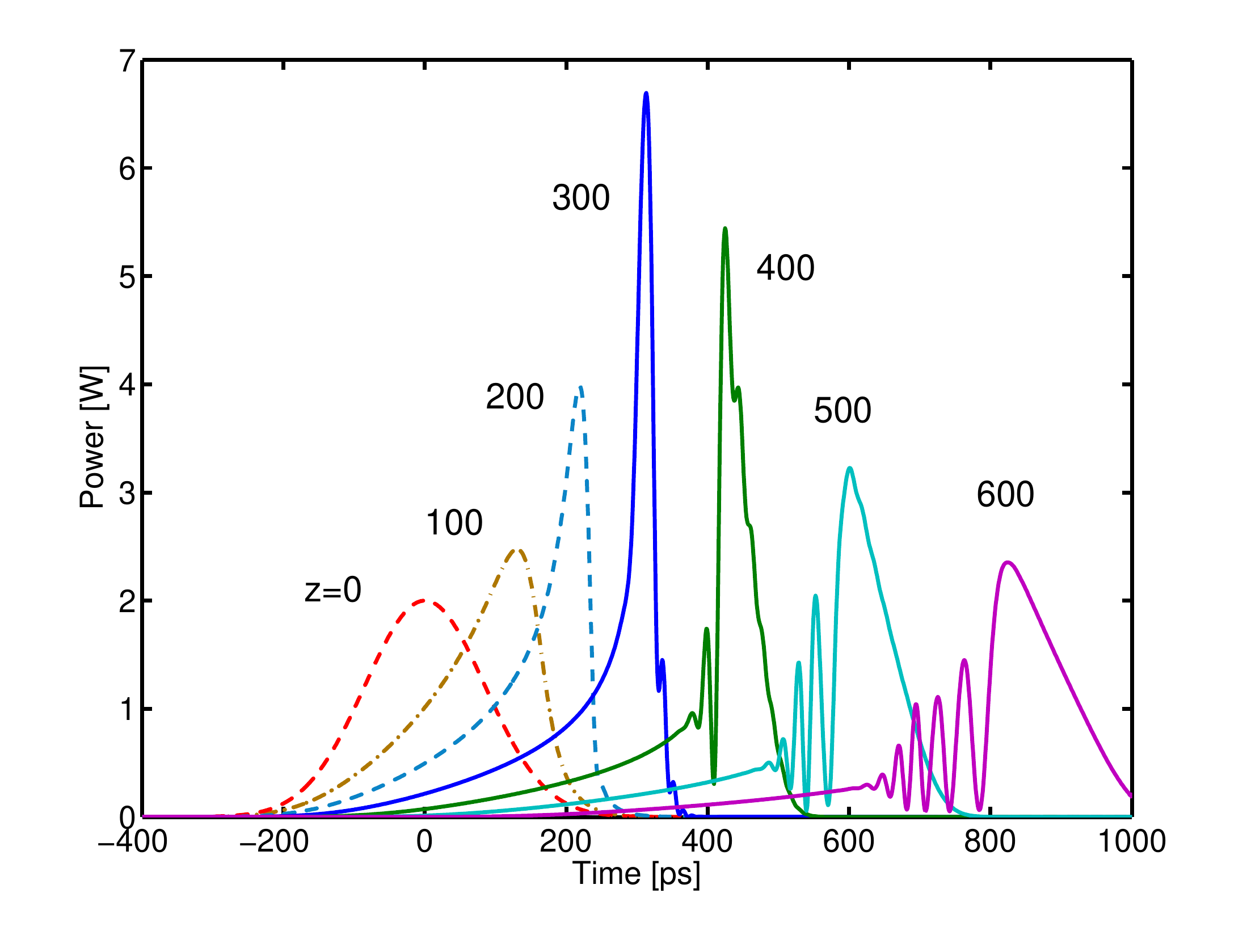}
\caption{Details of pulse evolution as in Fig.\ref{fig10}; numbers next to each curve indicate the propagation distance in meters.}
\label{fig11}
\end{figure}

A selection of pulse power profiles at a given set of distances (indicated in meters next to each curve) along the fiber, corresponding to the power contour plot of Fig.\ref{fig10}, is shown in Fig.\ref{fig11}. Here we can see that the formation of a vertical trailing edge at $z=300\: m$ is accompanied by a nearly three-fold amplification of the peak power value. Past the shoaling point, the peak power decreases, and a temporal oscillation appears in the leading edge while the trailing edge remains smooth.    

\begin{figure}[ht]
\centering
\includegraphics[width=10cm]{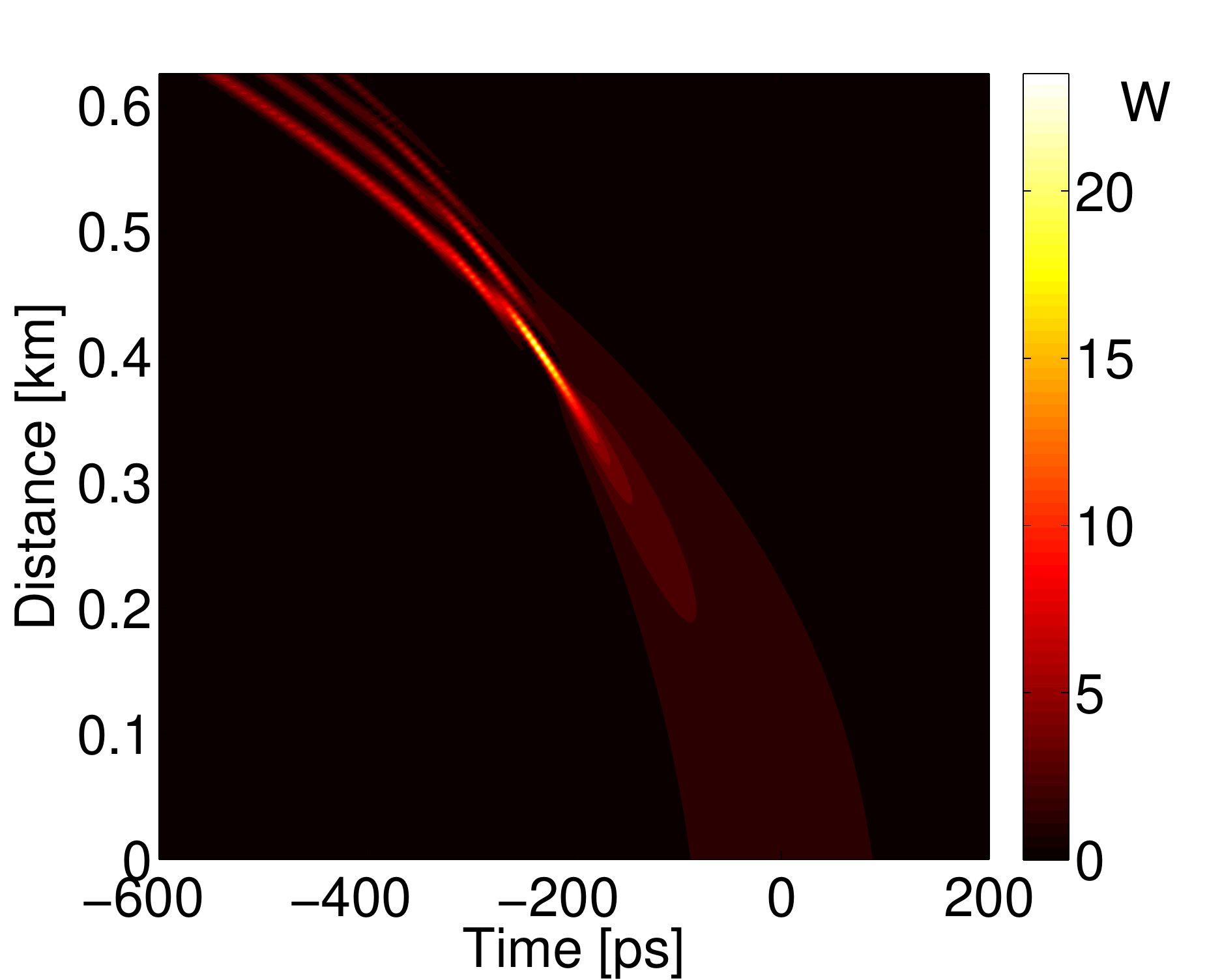}
\caption{Same as in Fig.\ref{fig10}, with positive third-order dispersion $\beta_3=2 \: ps^3/m$.}
\label{fig12}
\end{figure}

Quite suprisingly, numerical simulations reveal that whenever the sign of the TOD is changed from negative to positive (i.e., we set $\beta_3=2 \: ps^3/m$), that is the frequency-shifting pulse sees a progressively larger normal GVD as it propagates along the fiber, the temporal compression and peak power enhancement of the input Gaussian Riemann pulse is greatly enhanced at the shoaling point. 

\begin{figure}[ht]
\centering
\includegraphics[width=10cm]{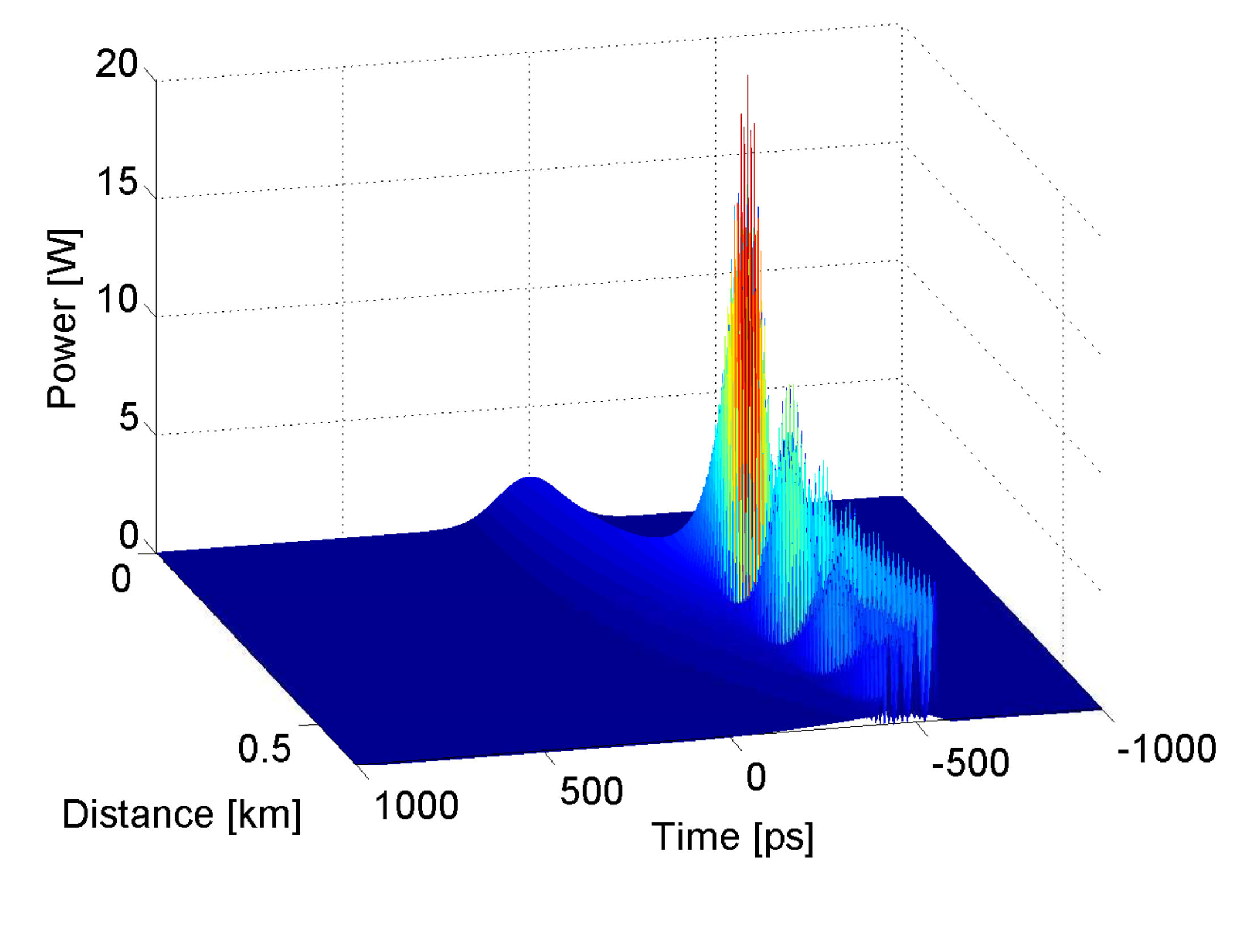}
\caption{Surface plot of pulse power vs. distance for the same case as in Fig.\ref{fig12}}
\label{fig13}
\end{figure}

In Figs.\ref{fig12}-\ref{fig13} we show the contour and surface plots of the pulse power evolution along the DCF for the same frequency up-shift rate as in Figs.\ref{fig10}, but the sign of TOD is changed from negative to positive. As it can be seen, in this case the vertical edge or shoaling occurs in the leading edge of the pulse at about $z=400\: m$. Moreover, the temporal compression and the pulse peak power is substantially increased with respect to the case of negative TOD which was illustrated in Figs.\ref{fig10}-\ref{fig11}. In particular, a more than ten-fold increase in the peak power is observed at the point of maximum temporal compression $z=400\: m$. 

\begin{figure}[ht]
\centering
\includegraphics[width=10cm]{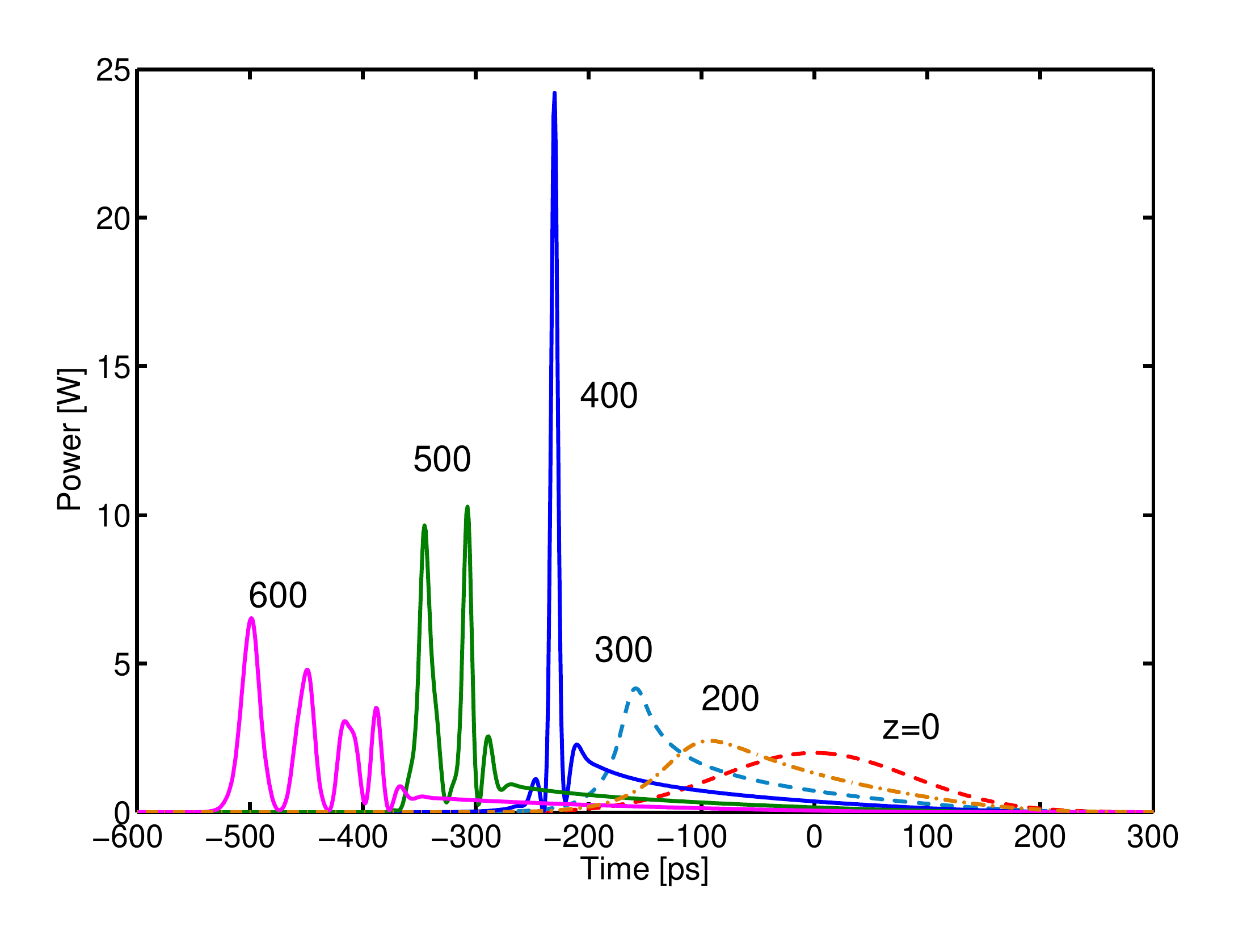}
\caption{Details of pulse evolution as in Fig.\ref{fig12}; numbers next to each curve indicate the propagation distance in meters.}
\label{fig14}
\end{figure}

A set of pulse power profiles for progressively increasing distances (indicated in meters next to each curve) in the fiber, extracted from the plots of Fig.\ref{fig12}-\ref{fig13}, is shown in Fig.\ref{fig14}. As it can be seen,  the compressed pulse reaches a peak power as high as $24\: W$ at $z=400\: m$. Past the point of maximum compression, the pulse break-up into multiple peaks until a pulse train is obtained at $z=600\: m$. 

\begin{figure}[ht]
\centering
\includegraphics[width=10cm]{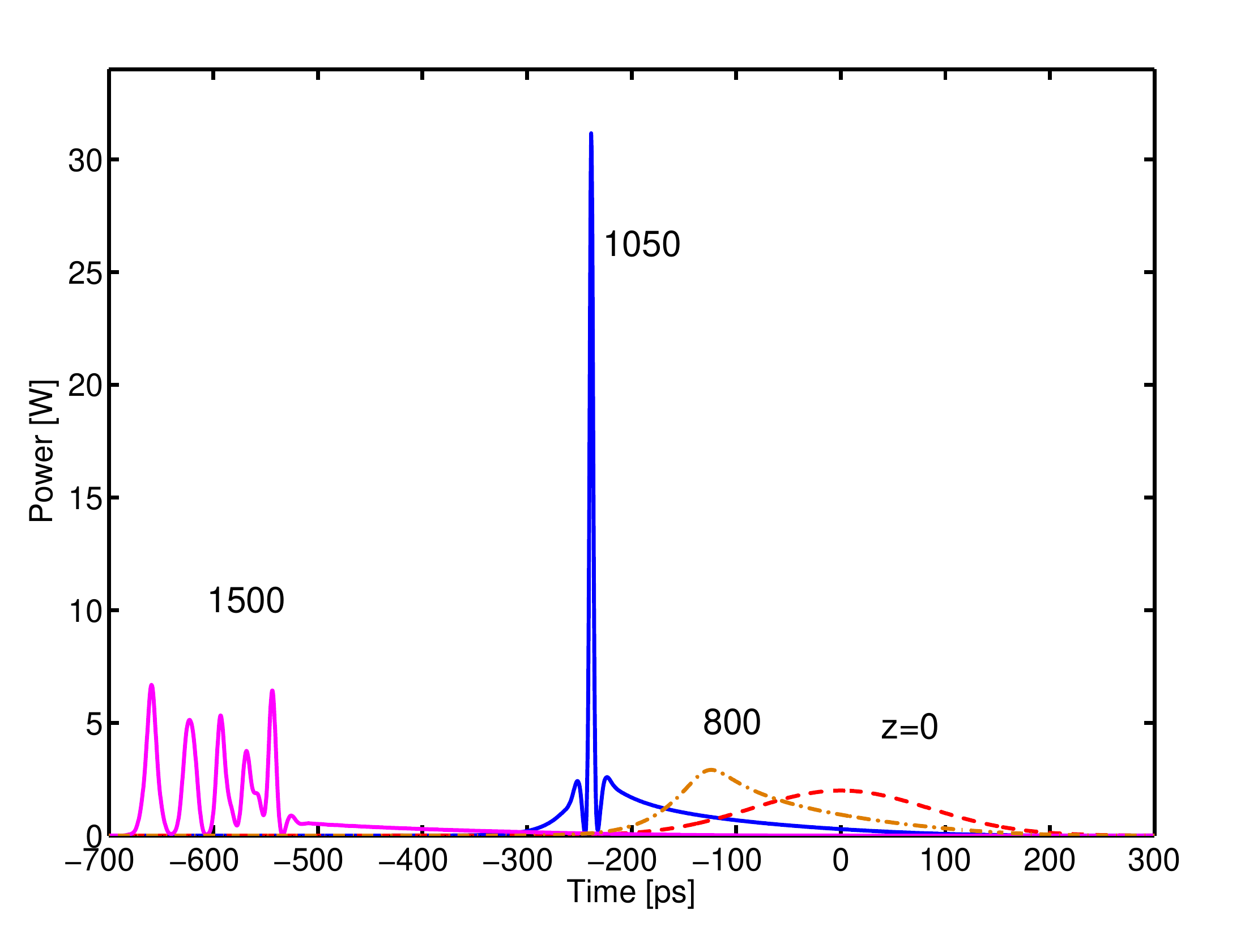}
\caption{Same as in Fig.\ref{fig14} with the positive third-order dispersion $\beta_3=0.4 \: ps^3/m$.}
\label{fig15}
\end{figure}

In the previous examples we have used a relatively large value of the TOD coefficient. However the dynamics of rogue wave formation is not critically dependent on the specific TOD value: in Fig.\ref{fig15} we show that a similar pulse compression and extreme peak formation also occurs whenever the TOD is reduced by five times to $\beta_3=0.4 \: ps^3/m$. The main difference between Fig.\ref{fig14} and Fig.\ref{fig15} is that the point of maximum compression is moved to further down the fiber, at $z=1\: km$ from $z=400\: m$. 

\begin{figure}[ht]
\centering
\includegraphics[width=10cm]{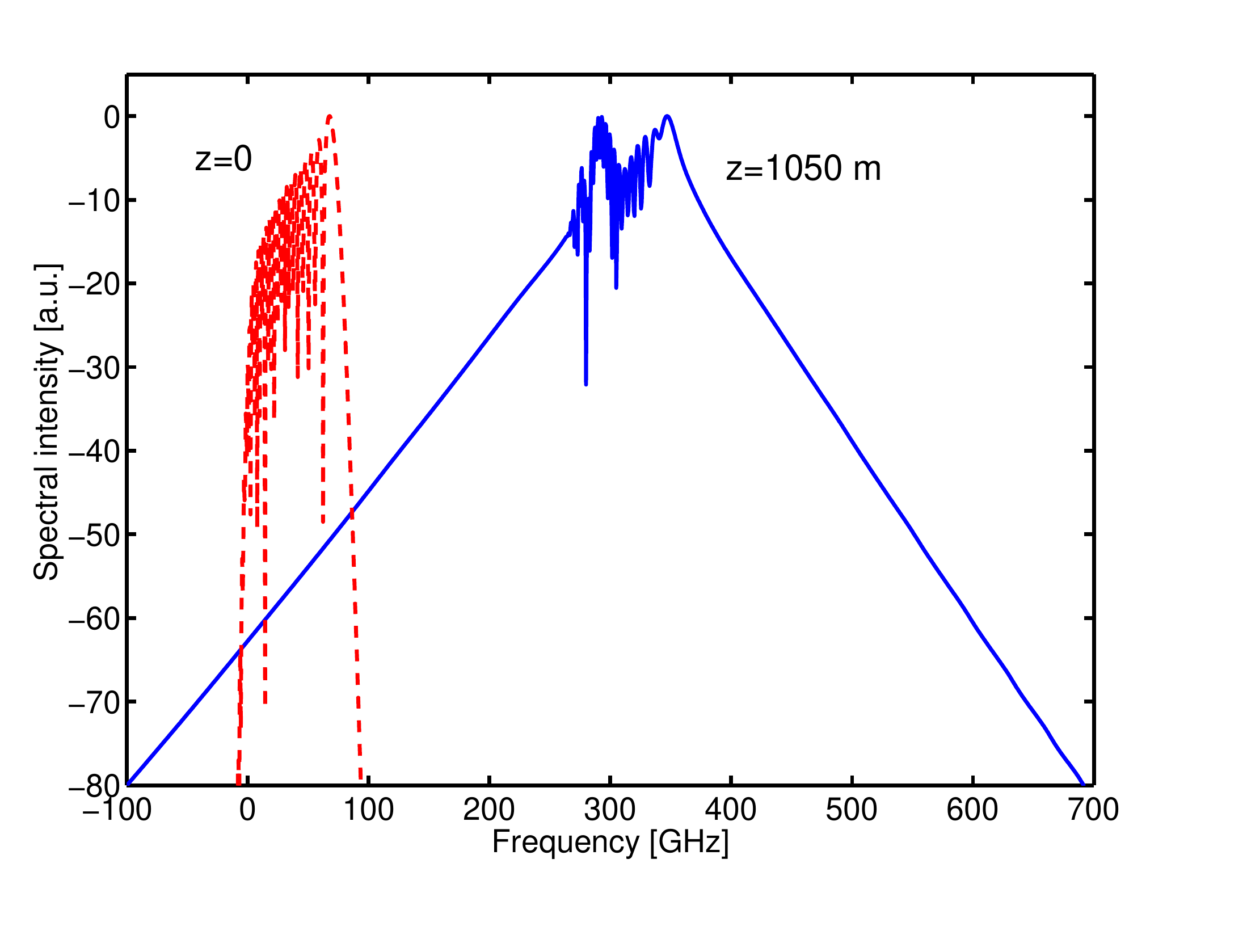}
\caption{Spectra of input pulse and of maximum compressed pulse as in Fig.\ref{fig15}.}
\label{fig16}
\end{figure}

Fig.\ref{fig16} displays the spectral intensities associated with the input pre-chirped pulse (red dashed curve) and the frequency up-shifted rogue pulse at $z=1\: km$ which are reported in Fig.\ref{fig15}. As can be seen, in spite of the relatively large normal GVD, the spectrum of the rogue pulse exhibits broad triangular tails that are associated with the presence of a sharp hyperbolic secant soliton-like peak. 

It is interesting to point out that, in contrast with the case of either chirp-free input pulses, which are subject to TOD-induced wave-breaking or shock formation in dispersion tapered fibers \cite{latkin07}-\cite{wab08}, in the case of Riemann pulses rogue wave formation is not accompanied by any wave-breaking phenomena. Note that whenever a step-wise pre-chirp of a CW is used, temporal shocks still arise even in the absence of TOD \cite{yuji99}-\cite{biondini06}. 

In order to highlight the nonlinear and dispersive mechanisms of pulse formation dynamics, in Eq.(\ref{nls1}) and elsewhere in this paper we neglected the presence of linear optical fiber loss. Indeed the effect of fiber loss is minimal and it does not qualitatively change our conclusions for the relatively short propagation lengths (up to a few km) which have been considered in our numerical examples. In general, effective transparent propagation can be achieved in optical fibers thanks to distributed Raman amplification, or by means of periodic lumped amplification in optical communication systems.     

\section{Conclusions}
\label{sec:concl}
In this work we have described optical pulse shoaling in the normal dispersion regime of optical fibers. We obtained exact solutions of the optical NSWE, and demonstrated their relatively good match with the numerical solutions of the NLSE. We have also revealed that TOD may lead to the occurrence of extreme waves or optical tsunamis in dispersion tapered fibers, in full analogy with the dramatic run-up and wave height amplification of ocean tsunamis. 

The present results may also have applications in a context different from nonlinear optics, such as hydrodynamics or BEC. Moreover, we envisage that optical Riemann waves and tsunamis may also be observed in the spatial domain, that is when diffraction replaces dispersion. In this case the presence of two-dimensional degrees of freedom may facilitate the generation and control of the necessary initial phase profiles \cite{wan}. 

\ack{}
The present research was supported in Brescia by Fondazione Cariplo, grant n.2011-0395.

\end{document}